%% file: sn-article_v1.tex
\documentclass[sn-nature]{sn-jnl}

\usepackage{threeparttable}  
\usepackage{multicol}  
\usepackage{multirow} 
\usepackage{graphicx}%
\usepackage{multirow}%
\usepackage{amsmath,amssymb,amsfonts}%
\usepackage{amsthm}%
\usepackage{mathrsfs}%
\usepackage[title]{appendix}%
\usepackage{xcolor}%
\usepackage{textcomp}%
\usepackage{manyfoot}%
\usepackage{booktabs}%
\usepackage{algorithm}%
\usepackage{algorithmicx}%
\usepackage{algpseudocode}%
\usepackage{listings}%
\usepackage{fancyhdr}
\usepackage{url}

\theoremstyle{thmstyleone}%
\newtheorem{theorem}{Theorem}
\theoremstyle{thmstyletwo}%

\theoremstyle{thmstylethree}%

\begin{document}

\title[Article Title]{Capturing Opportunity Costs of Batteries with a Staircase Supply-Demand Function}

\author*[1]{\fnm{Ye} \sur{Guo}}\email{guo-ye@sz.tsinghua.edu.cn}
\author[1]{\fnm{Chenge} \sur{Gao}}\email{gcg23@mails.tsinghua.edu.cn}
\author[2]{\fnm{Cong} \sur{Chen}}\email{cc2662@cornell.edu}

\affil*[1]{\orgdiv{Tsinghua Berkeley Shenzhen Institute}, \orgname{Tsinghua University}, \orgaddress{\street{Lishui Road}, \city{Shenzhen}, \postcode{518000}, \state{Guangdong}, \country{China}}}
\affil[2]{\orgdiv{School of Electrical and Computer Engineering}, \orgname{Cornell University}, \orgaddress{\street{Hoy Road}, \city{Ithaca}, \postcode{14850}, \state{New York}, \country{USA}}}


\abstract{In the global pursuit of carbon neutrality, the role of batteries is indispensable. They provide pivotal flexibilities to counter uncertainties from renewables, preferably by participating in electricity markets. Unlike thermal generators, however, the dominant type of cost for batteries is opportunity cost, which is more vague and challenging to represent through bids in stipulated formats. This article shows the opposite yet surprising results: The demand-supply function of an ideal battery, considering its opportunity cost, is a staircase function with no more than five segments, which is a perfect match with existing rules in many real electricity markets. The demand-supply function shifts horizontally with price forecasts and vertically with the initial SOC. These results can be generalized to imperfect batteries and numerous battery-like resources, including battery clusters, air-conditioners, and electric vehicle charging stations, although the number of segments may vary. These results pave the way for batteries to participate in electricity markets. }




\maketitle

\section{Introduction}\label{sec1}
\subsection{Background}

\input Introduction_v1


\input RelatedWork_v1

\begin{figure}[H]
	\centering
        \vspace{-1cm}
	\includegraphics[width=5in]{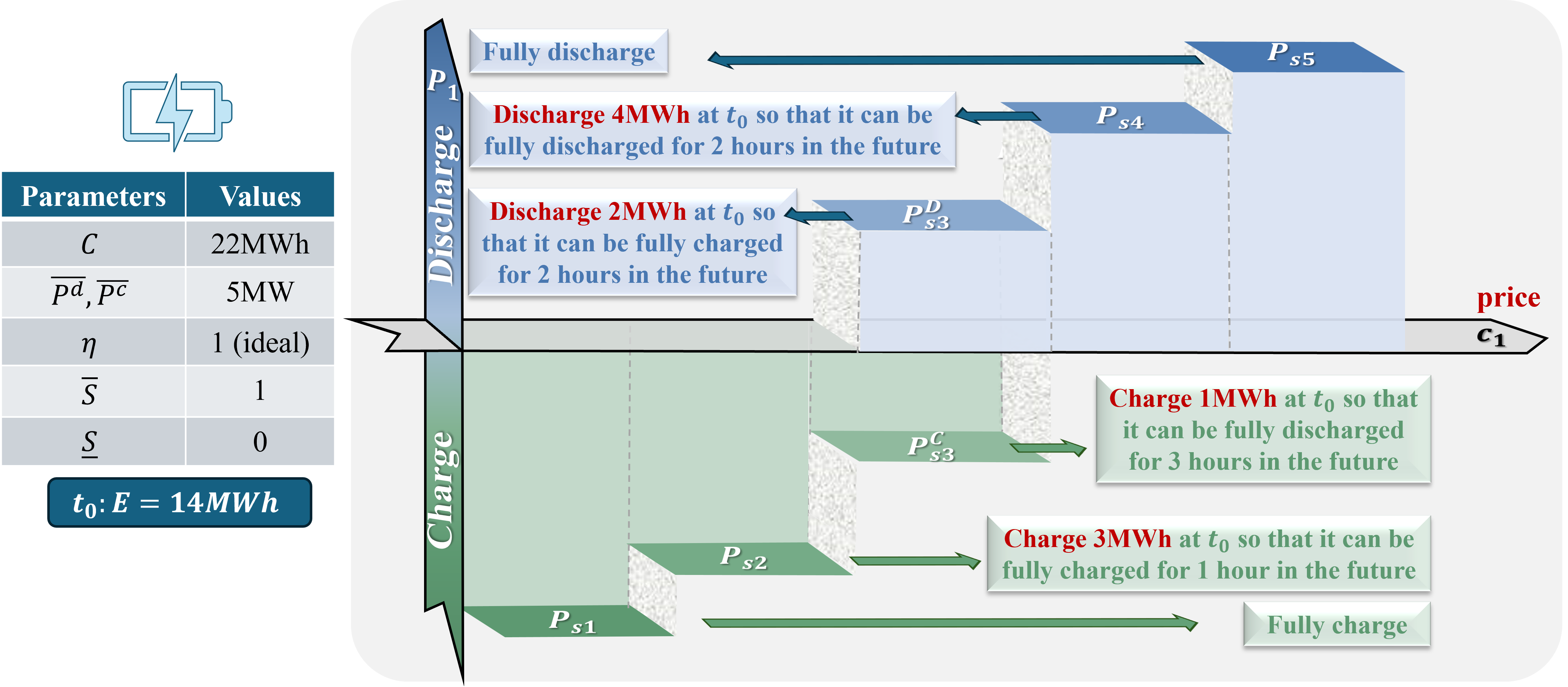}
	\caption{An example of staircase demand-supply function of a lossless battery with in maximum of five stairs. (There are six possible cases but $P^D_{s3}$ and $P^C_{s3}$ will not exist simultaneously) 
 } 
	\label{fig:5seg}
\end{figure}

In this paper, we prove a succinct but interesting result: Under mild assumptions, the opportunity cost of an ideal battery can be represented by a staircase demand-supply function with no more than five segments. This is a coincidental but perfect match with current rules in many regional electricity markets. Moreover, each possible stair has a clear meaning, as illustrated in Fig. \ref{fig:5seg}. Considering charging and discharging inefficiencies, energy dissipation, and ending SOC limits, we may see more stairs, but the shape of a staircase function remains. Observations of batteries' discrete actions are consistent with some existing works, e.g., \cite{Lamadrid24JRE, ZhengXu22socAribitrage}. Theoretical proofs are also given in this paper together with these observations.

It is also shown that as price forecasts change, demand-supply functions shift purely horizontally, while with a change of initial SOC, they shift purely vertically. In addition, it should be noted that many conclusions above also hold for a wide range of battery-like resources, such as storage stations, charging stations, and flexible loads of air conditioners. Results presented in this paper pave the way for all such resources to actively participate in the electricity market and for the market administrator to monitor their bids, which are of crucial importance in our transitions to clean energy and a sustainable world.

\section{Assumptions and Basic Results}\label{sec2}

The following assumptions are made for a battery throughout the paper:

i) The battery operator considers the current and multiple future time intervals and bids its demand-supply function, which formulates its charging/discharging power as a function of price for the current interval.

ii) The demand-supply function above is obtained based on the profit maximization problem of the battery, considering future price forecasts and its capacity limits and solving for charging/discharging levels with varying prices for the current time interval.

We start with the simplest case of an ideal battery, which has 100\% charging and discharging efficiencies, no energy dissipation, and equal limits on charging and discharging electric power. With assumptions above, the demand-supply curve of such an ideal battery is a staircase function with no more than five segments, each has a clear physical background, 

Specifically, there are six possible stairs shown in Fig.2, each with a clear physical background. Without loss of generality, assume the length of each period is one hour and that the battery needs $n_C$ hours to charge from its initial SOC to the upper bound and discharge $n_D$ hours to discharge from its initial SOC to the lower bound. In general, $n_C$ and $n_D$ may not be integers. Then the six possible stairs are explained in sequence as follows: 

\begin{enumerate}
    \item Fully charge
    \item Charge-for-charge: Charge remainder energy to charge for $\lfloor n_C\rfloor$ hours fully
    \item Charge-for-discharge: Charge remainder energy to discharge for $\lceil n_D\rceil$ hours fully 
    \item Discharge-for-charge: Discharge remainder energy to charge for $\lfloor n_C\rfloor$ hours fully
    \item Discharge-for-discharge: Discharge remainder energy to discharge for $\lceil n_D\rceil$ hours  fully
    \item Fully discharge
\end{enumerate}

\begin{figure}[H]
	\centering
        \vspace{-1cm}
	\includegraphics[width=5.0in]{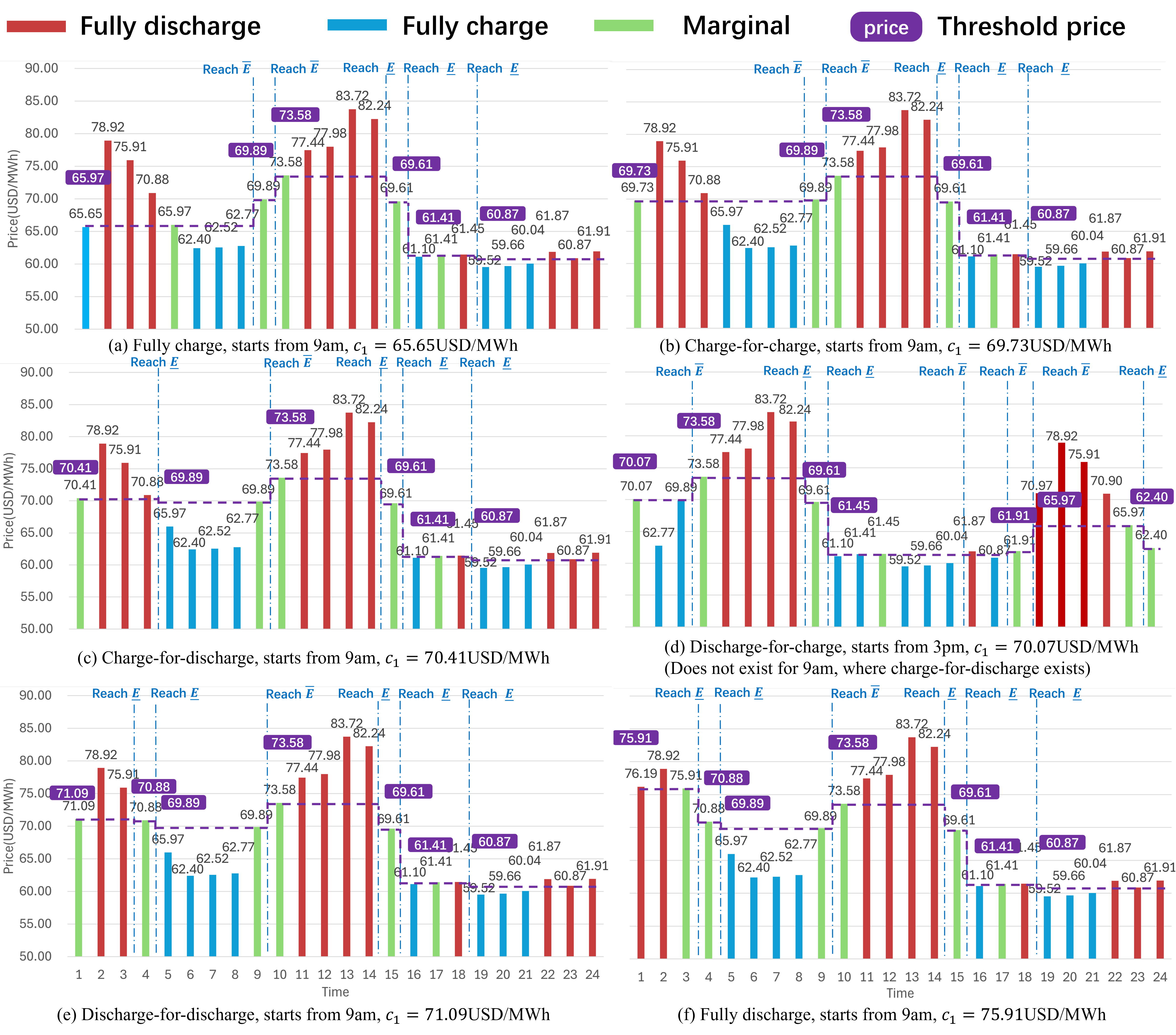}
	\caption{Illustrations of action sequence of all six possible segments. The battery is same as Figure \ref{fig:5seg}. Prices are cycling day-ahead LMPs of PJM on May.8, 2022, at pnode\_id=49722 (in PEPCO) \cite{PJMLMPs}). Prices of all periods are compared to a threshold (purple), higher leads to fully discharge, lower leads to fully charge, and equal may lead to marginal (neither of former two). The threshold is initialized as $c_1$ if it is marginal, as in (b)-(e), and changes when energy limits are reached.
 } 
	\label{fig:figz}
\end{figure}

It should be noted that the number of actual segments we see maybe even smaller than five. Please see Fig.\ref{fig:figx}(d-f) as examples.

\section{Imperfect Batteries}

We generalize results in the previous section to imperfect batteries whose efficiencies are smaller than 100\% and with energy dissipation. When studying imperfect batteries, the following assumption is further added:

iii) Electricity prices for all periods are non-negative.

\begin{figure}[H]
	\centering
	\includegraphics[width=5.0in]{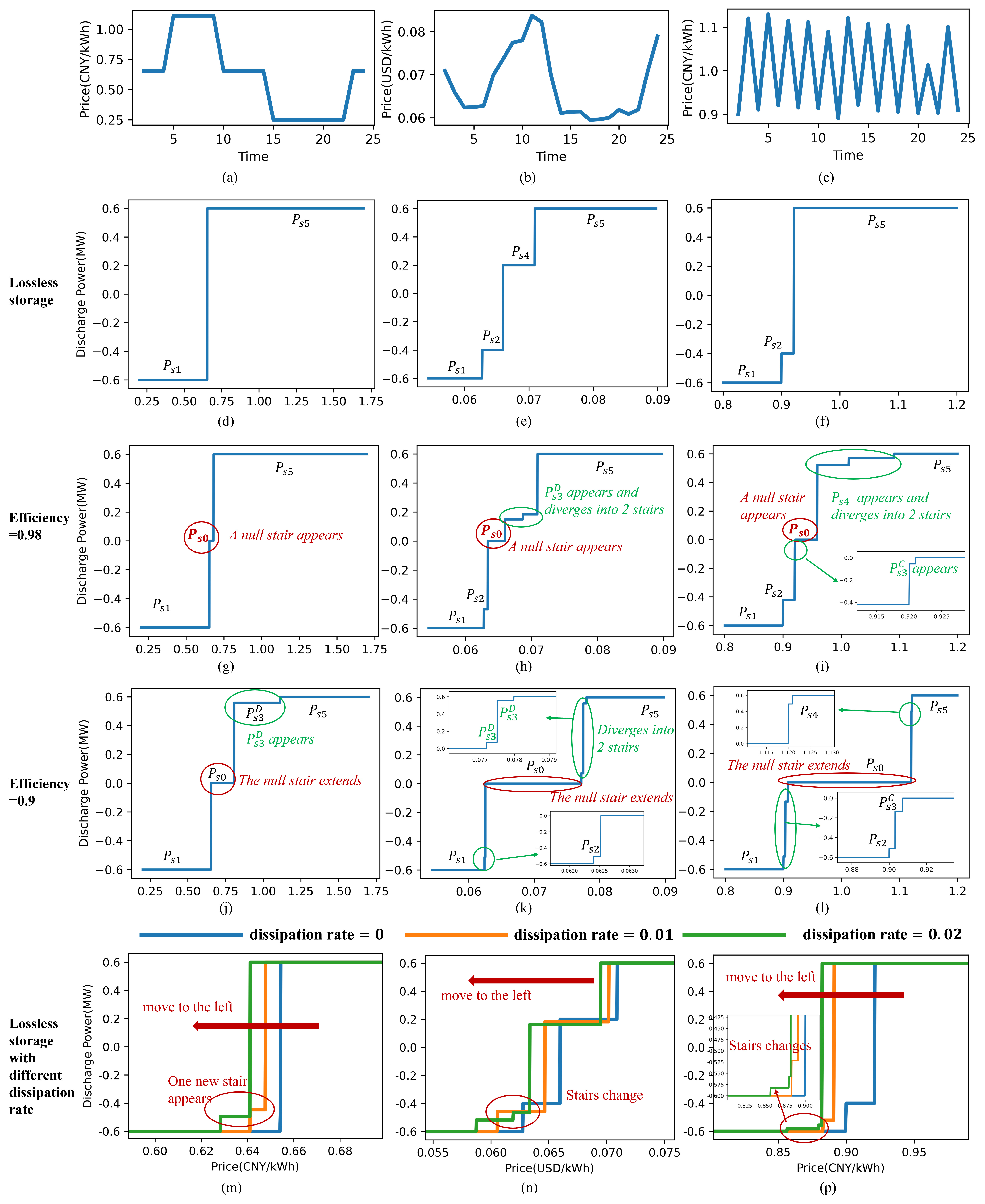}
        \caption{\small Results of the demand-supply function of the energy storage system (the capacity is 2MWh, the maximum charging/discharging power is 0.6MW, the upper and lower bounds of SOC are 1 and 0.1, the initial SOC is 0.5, the total number of periods considered is 24 with each period being 1 hour). Sub-figures (a)-(c) are three settings of price forecasts, from time-of-use prices in Shenzhen \cite{SZpricedata} updated in 2022, the day-ahead LMPs of PJM (May.8, 2022, pnode\_id=49722, which is a 13kV LOAD node in zone PEPCO \cite{PJMLMPs}), and artificially set extremely fluctuating prices. Sub-figures (d)-(f) are demand-supply functions of an ideal battery with the parameters above, with two, four, and three stairs, respectively. Sub-figures (g)-(i) are demand-supply functions of a battery with charging and discharging efficiencies of 0.98, where the null segment appears, and stair divergence can be observed in (h) and (I). Sub-figures (j)-(l) are demand-supply functions of a battery with charging and discharging efficiencies of 0.9, where null stairs extend, indicating the battery is less motivated to act due to possibly significant power loss. Sub-figures (m)-(p) shows the impact of the \textbf{energy dissipation} of a storage system to the demand-supply function. It can be observed that when the energy dissipation increases, the demand-supply function will move to the left, with the values of stairs changing. Also, existing stairs may disappear and new stairs may appear when the energy dissipation increases.}

	\label{fig:figx}
\end{figure}

Examples of demand-supply curves are given in Fig.\ref{fig:figx}(g-l) with imperfect charging/discharging efficiencies, and Fig.\ref{fig:figx}(m-p) for examples with energy dissipation.

Compared to an ideal battery, the results above introduce two notable differences. In addition to the proof provided, we also give intuitive explanations as follows:

\begin{enumerate}
    \item An additional null stair may appear: When charging and discharging introduce energy loss, batteries will become reluctant to act when the anticipated price difference in the future is not large enough. With increasing energy losses caused by charging, discharging, or energy dissipation, the null stair tends to become more common and wide, which can also be observed in Fig.\ref{fig:figx}(g-l) and Fig.\ref{fig:figx}(m-p).
    \item Some stairs in the ideal case may diverge to multiple ones: The fundamental reason is that varying current-interval prices in the x-axis may lead to different actions in upcoming intervals, and different energy loss associated may differentiate the value of the ``remainder energy''. Take the Charge-for-charge segment as an example, different actions in future intervals, e.g., charging remainder-fully charge versus charging remainder-fully discharge-fully charge-fully charge, actually define different values of the remainder energy considering energy loss, which explains the possible divergence of stairs.
\end{enumerate}

\newpage

\section{Possibly Negative Prices}

\begin{figure}[H]
	\centering
	\includegraphics[width=5.0in]{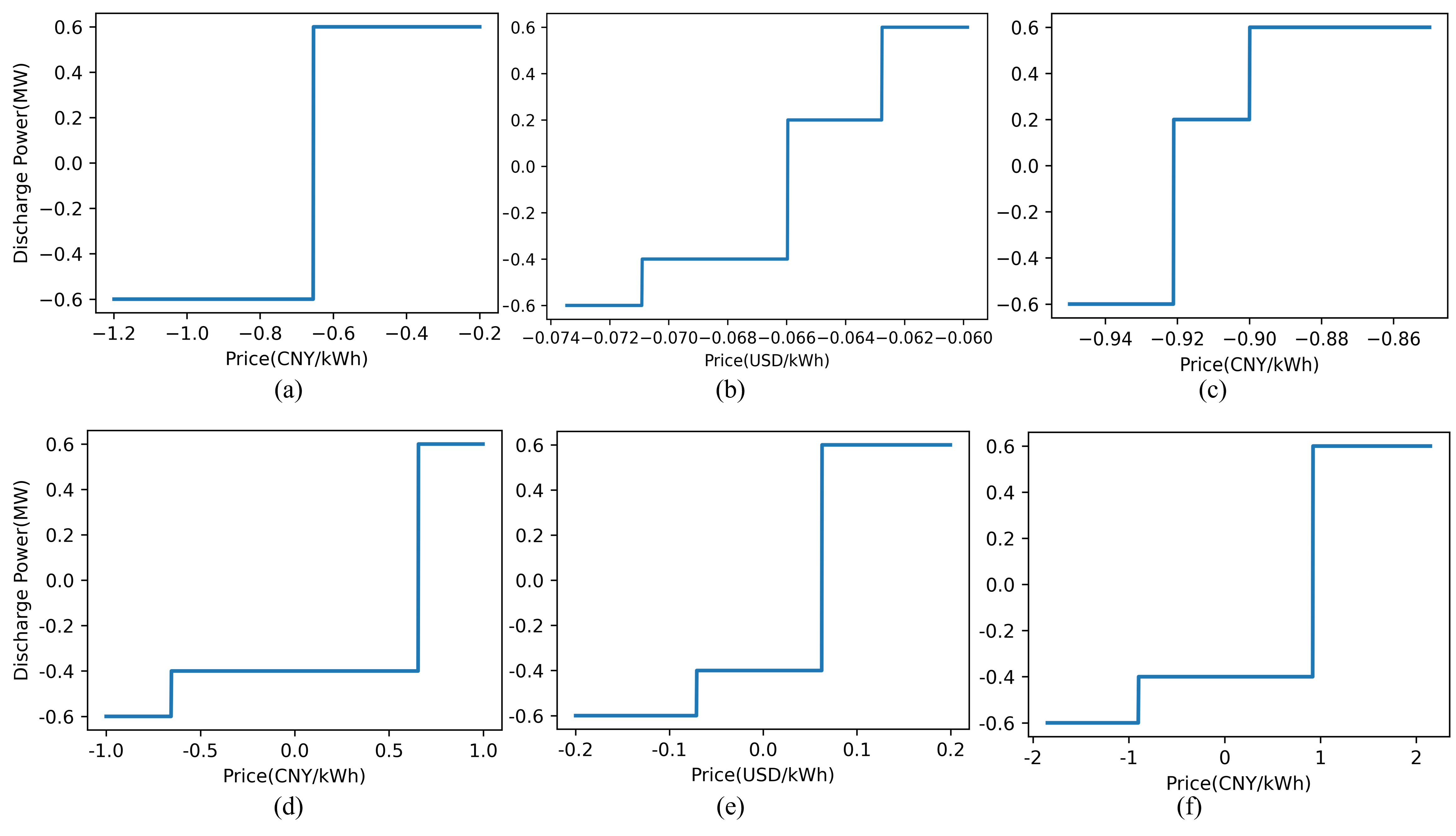}
	\caption{Impact of \textbf{negative electricity prices} of a storage system to the demand-supply function (the capacity is 2MWh, the maximum charging/discharging power is 0.6MW, the upper and lower bounds of SOC are 1 and 0.1, the efficiencies are 1. Subfigures (a), (b) and (c) show the demand-supply functions under constant negative prices (adding minus signs before Shenzhen, PJM, and extreme prices for all periods, respectively). Subfigures (d), (e) and (f) show those under scarce negative prices (adding minus signs for the second period only). It can be observed that when negative prices occur, the demand-supply function is still staircase and monotonically non-decreasing.}
	\label{fig:negativeprice}
\end{figure}

There are more and more frequent reports of negative prices, which is one of our concerns about assumption iii). Although our proof relies on such an assumption, as is presented in the supplemental file, it is observed that conclusions in the previous section still hold with both scarce and long-lasting negative prices, as illustrated in Fig. \ref{fig:negativeprice}.

\section{Parameter Sensitivities}

\begin{figure}[h!]
	\centering
	\includegraphics[width=5.0in]{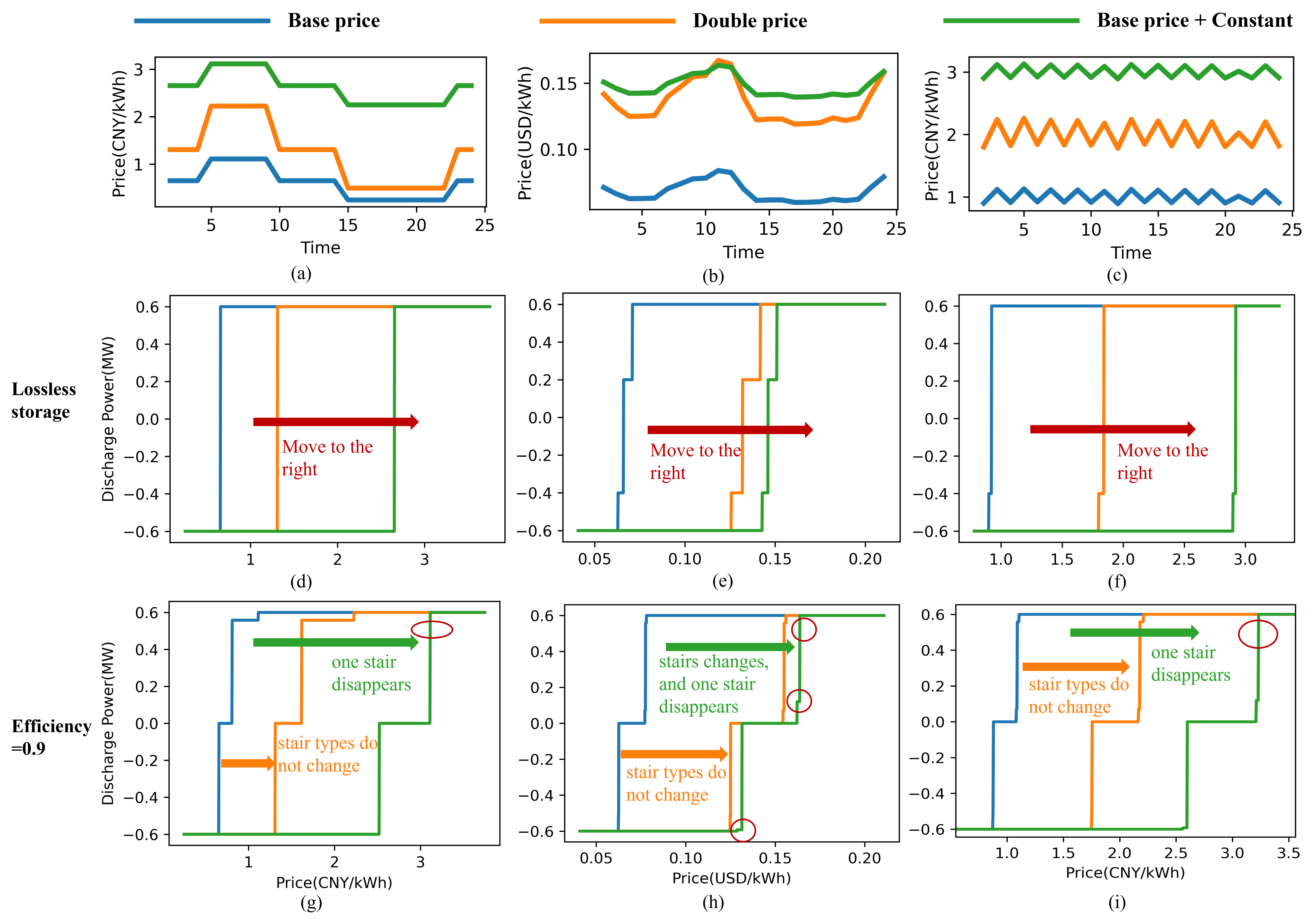}
	\caption{Impact of \textbf{future prices} on demand-supply functions of the storage system(the capacity is 2MWh, the maximum charging/discharging power is 0.6MW, the upper and lower bounds of SOC are 100\% and 10 \%, the efficiencies are 1(for (d), (e), and (f)) and 0.9(for (g), (h), and (i)), the total number of time periods is 24 with one time period being 1 hour). (a), (b) and (c) show electricity prices in future $T-1$ time periods. Constants for (a), (b), and (c) are 2CNY/kWh, 0.08USD/kWh, and 2CNY/kWh. In (d), (e) and (f), it can be observed that if future prices increase, functions will move to the right. In (g), (h), and (i), it can be observed that if future prices change in an equal proportion, the demand-supply function will move horizontally, and the types of stairs will not change; but if future prices change in an unequal proportion, ratios of prices between different periods will change, which results in stair types changing.}
	\label{fig:pricesen1}
\end{figure}

\textbf{Impact of future price predictions:} When future price forecasts change with other parameters unchanged, the demand-supply function of a battery shifts in a horizontal direction. In particular, if future prices change proportionally, e.g., with different currencies, the shape of the curve remains the same. Otherwise, its shape may differ from the original one, with each segment moving in different magnitudes, possibly leading to more or less stairs. Examples are given in Fig.\ref{fig:pricesen1}. 

\begin{figure}[h!]
	\centering
	\includegraphics[width=5.0in]{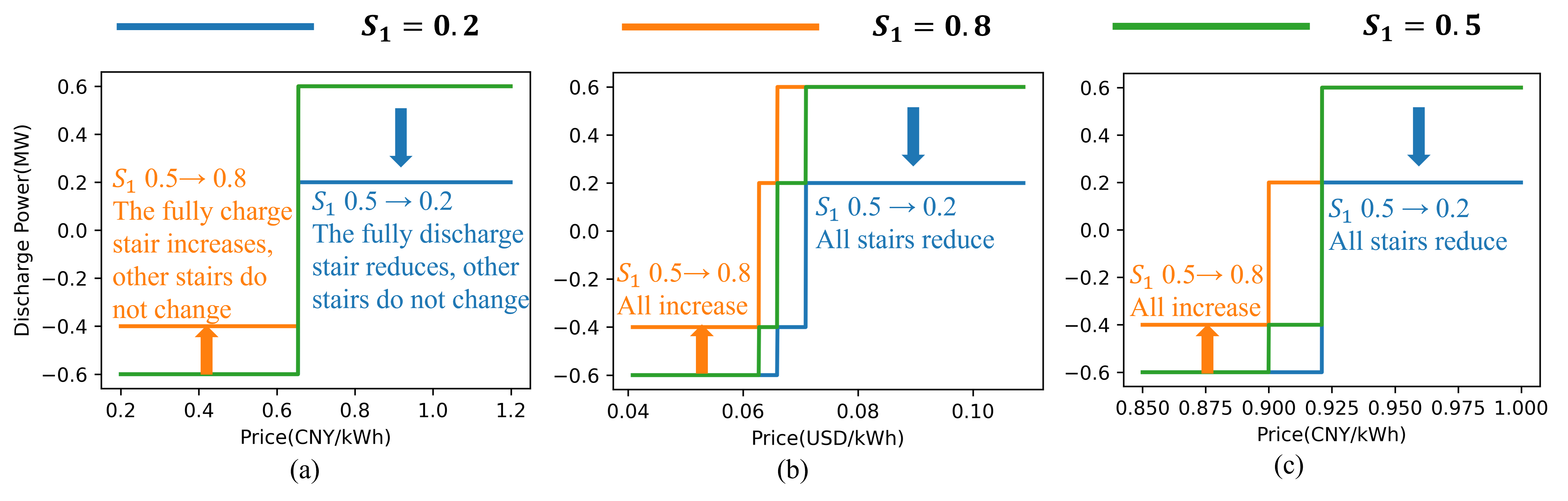}
	\caption{Impact of \textbf{the initial SOC} of a lossless storage system on the demand-supply function (the capacity is 2MWh, the maximum charging/discharging power is 0.6MWh, the upper and lower bounds of SOC are 100\% and 10\%. (a), (b) and (c) show results of demand-supply functions under prices of Shenzhen, PJM, and extreme prices respectively (Fig.\ref{fig:figx}(a), (b), and (c)). It can be observed that when the initial SOC increases, the demand-supply function function will first move upward until touches the fully discharge stair.}
	\label{fig:E1change}
\end{figure}

\textbf{Impact of Initial SOC:} If the initial SOC changes while other parameters do not, the demand-supply function will move vertically, as shown in Fig.\ref{fig:E1change}.

\section{Generalized ESS resources}

\subsection{ESS cluster}
\par The demand-supply function of an ESS cluster can be calculated by summing up all functions of individual ESSs. 
\begin{figure}[h!]
	\centering
	\includegraphics[width=5.0in]{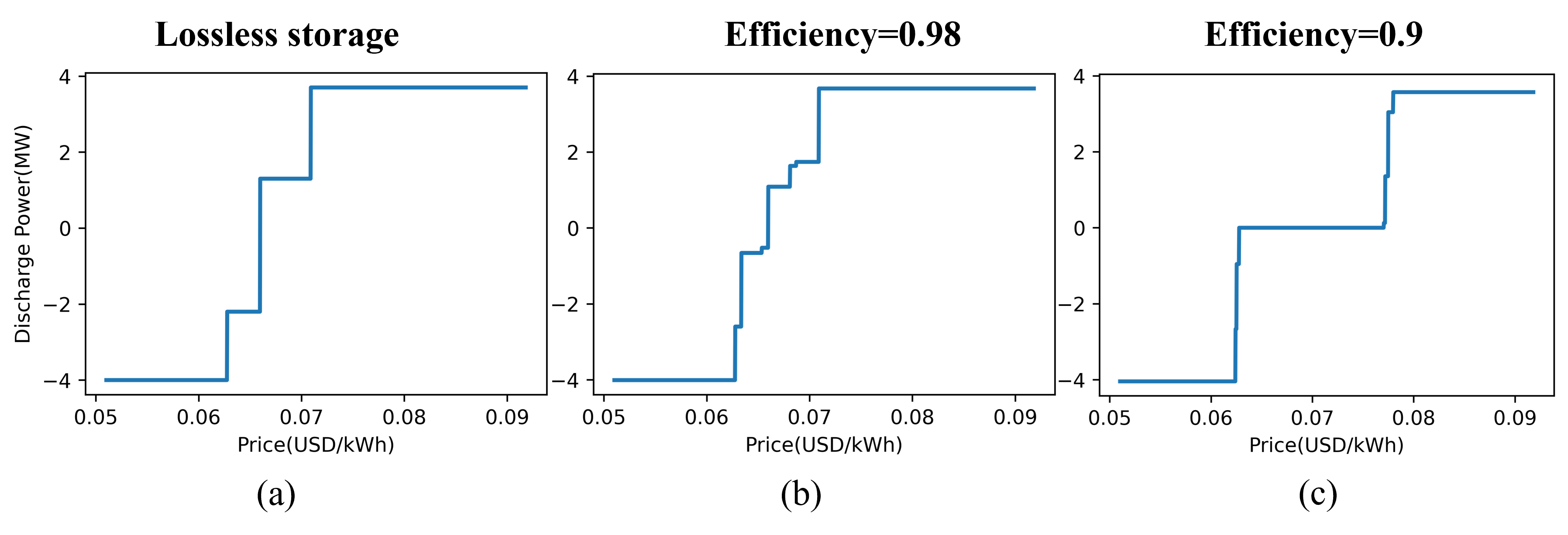}
	\caption{The demand-supply function of an \textbf{ESS cluster} (for all seven storage units in the cluster, the capacity is 2MWh, the maximum charging/discharging power is 0.6MW, the upper and lower bounds of SOC are 1 and 0.1; the initial SOCs are set as 70, 50, 20, 40, 80, 60, and 35\%) under electricity prices of PJM (Fig.\ref{fig:figx}(b)). It can be observed that the demand-supply function functions are staircase.}
	\label{fig:ESScluster}
\end{figure}
\subsection{Air conditioning}
\par For air conditioning (AC), here we only consider cooling. The demand-supply function of an AC system is a staircase function with no more than four kinds of segments. The proofs can be found in the Appendix. 
\begin{figure}[htbp]
	\centering
	\includegraphics[width=5.0in]{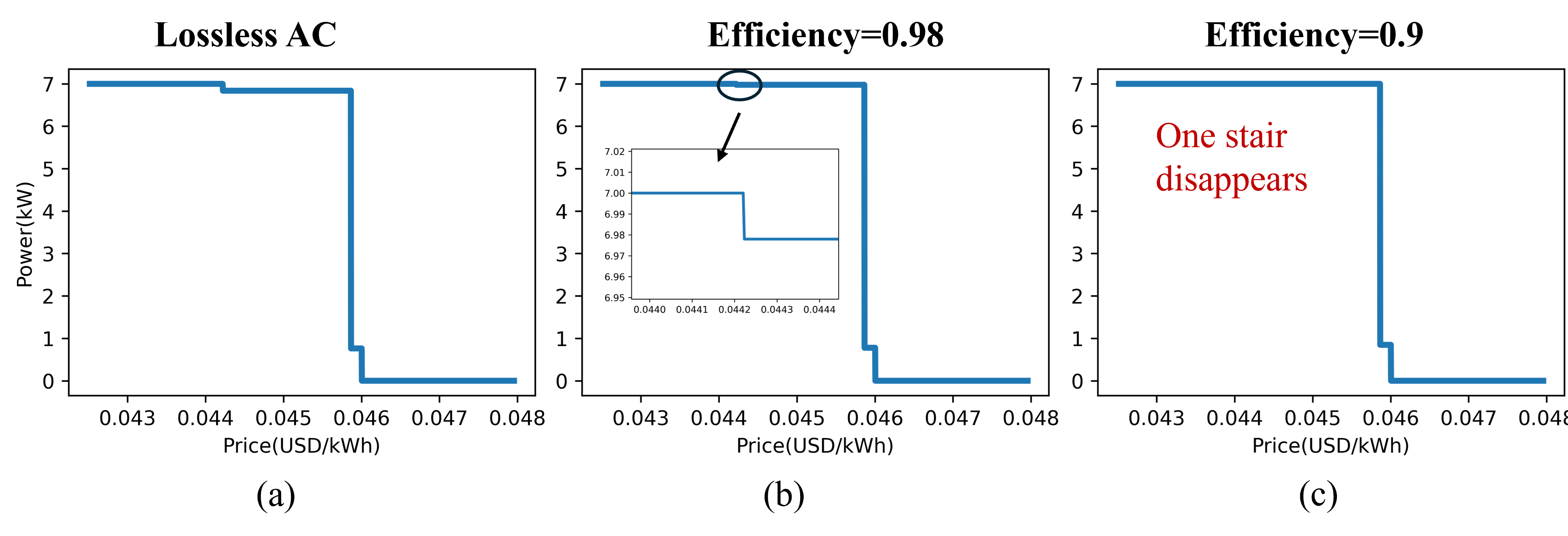}
	\caption{The demand-supply function of an \textbf{AC system} under electricity prices of PJM (Fig.\ref{fig:figx}(b)). Here, the AC system is modeled by a 1R-1C equivalent model, and only the cooling process is considered. The maximum power output, equivalent heat capacity, and equivalent thermal resistance are set to 7kW, 5$kW/^\circ C$, and 10$^\circ C/kW$ respectively. The initial room temperature and the upper and lower bounds of room temperature are set to 24$^\circ C$, 26$^\circ C$, and 22$^\circ C$ respectively. It can be observed that the demand-supply function function is staircase.}
	\label{fig:AC}
\end{figure}

\subsection{EV aggregator}
\par For electric vehicles (EVs), here we consider V2G. The demand-supply function of an EV is a staircase function. 
\begin{figure}[h!]
	\centering
	\includegraphics[width=5.0in]{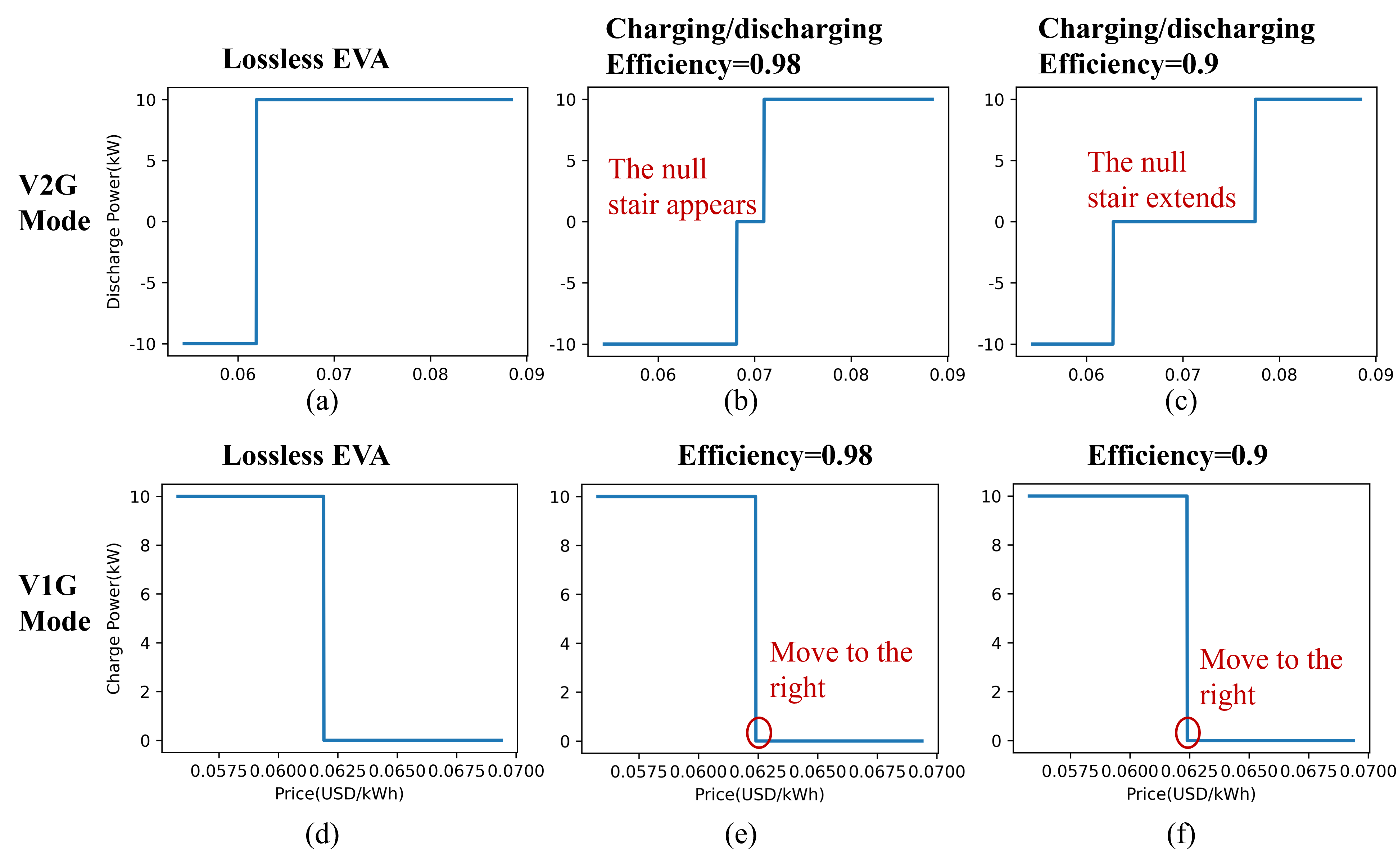}
	\caption{The demand-supply function of an \textbf{EVA} (there are 90 EVs in the EVA. For each EV, the capacity is 45kWh, the maximum charging/discharging power is 10kW; the arrival time is in a normal distribution with values of expectation and variance being 7:30 and 2.5 hours; the departure time is in a normal distribution with values of expectation and variance being 18:00 and 2 hours; the arrival SOC is in a uniform distribution within [0.2,0.4]; the departure SOC is in a uniform distribution within [0.8,1]) under electricity prices of PJM (Fig.\ref{fig:figx}(b)). It can be observed that the demand-supply function function is staircase.}
	\label{fig:EVcluster}
\end{figure}

\section{Conclusions and Discussions}

The opportunity cost of an ideal battery is a staircase function with no more than five segments. This is a fortuitous and perfect consistency between the physical backgrounds of batterys' operation problems and artificial electricity market rules. However, as we show later, no real battery is perfect, and imperfect batteries' demand-supply functions may deviate from the five-segment form, with an additional null segment and possible bifurcations. The primary objective of this article is to reveal the truth and establish properties. How this would affect the electricity market is an interesting problem that is worth further investigation.

\newpage
\input method_v1.tex

\backmatter

{
\bibliography{Ref}
}

\end{document}

%% file: Introduction_v1.tex
\par As an urgent and indispensable measure to achieving a sustainable world, the goal of carbon neutrality has become a global consensus. Power generation is the largest source of carbon emission by sector globally \cite{nationalacademies,mckinsey}. In China, which is the country with the largest amount of carbon emission, it accounts for more than 40\% \cite{Chinaecologyandenvironment}. Thus the transition from fossil-fuelled power generation to renewable energy sources (RES) is a paramount mission in our efforts towards a carbon-neutral and sustainable society.

The rapid development of renewable energy over recent years has been transformative. According to the International Energy Agency (IEA), global renewable energy capacity is projected to reach approximately 7,300 gigawatts (GW) by 2028———an expansion of nearly 2.5 times the current capacity. By 2025, renewables are expected to surpass coal, becoming the largest source of global electricity generation \cite{iearenewables}. These well-developed systems are poised to play a critical role in reducing carbon footprints and advancing global climate goals \cite{verpoort2024impact}, while also delivering significant economic benefits, including lower energy costs and reduced exposure to fossil fuel price volatility \cite{rode2021estimating}.

However, unlike conventional thermal units, wind turbines and photovoltaic (PV) panels lack inventories of primary energy, making the management of their inherent variability one of the most pressing challenges for contemporary power systems \cite{millstein2021solar, USeia}. At higher levels of renewable penetration, there are instances where wind and solar output exceeds grid demand, leading to renewable curtailment. Curtailment rates can escalate sharply with increased PV penetration, potentially surpassing 50\% in high PV penetration scenarios \cite{frew2021curtailment}, which poses a significant threat to the economic viability of these renewable sources \cite{heptonstall2021systematic}.

Addressing these challenges necessitates the deployment of advanced energy storage solutions, with batteries playing a pivotal role. Short-term storage (up to 12 hours) is effectively managed by lithium-ion batteries and pumped hydropower storage (PHS), known for their cost-effectiveness and reliability. By 2040, advancements in lithium-ion technology are expected to reduce energy consumption by 62–71\%, depending on cell chemistry \cite{degen2023energy}. For mid-duration needs, vanadium redox flow batteries (VRBs) strike a balance between cost and performance, while long-duration energy storage (LDES) (beyond 36 hours) is best served by hydrogen systems—particularly those with geologic storage—and natural gas with carbon capture (NG-CCS) \cite{hunter2021techno}. In systems where renewables contribute 60–90\% of electricity, LDES is essential, potentially cutting system costs by up to 50\% when energy capacity costs fall below 20USD/kWh \cite{sepulveda2021design, dowling2020role}. The advancement of these technologies is crucial for large-scale battery integration, enabling peak load shaving, reduced renewable intermittency, and improved regulation services \cite{wang2023accelerating}. 

As more batteries are deployed in the electric power system, they should naturally become active participants in the electricity market, similar to other types of resources \cite{ferc}. Given their charging and discharging capabilities, batteries may bid a demand-supply function, representing their willingness to buy or sell at different prices. In the ideal case, we expect batteries to bid truthfully, bidding their supply-demand functions according to their costs.

%% file: RelatedWork_v1.tex
Batteries' intrinsic costs may consist of multiple components: fixed costs \cite{peng2023heterogeneous,comello2019emergence}, degradation costs \cite{he2018intertemporal,levin2023energy}, etc. In the context of real-time or hourly operation, the dominant component is their opportunity costs: Charging power costs money, but it earns opportunities to sell back in the future, and vice versa. 
\begin{figure}[h]
	\centering
	\includegraphics[width=5.0in]{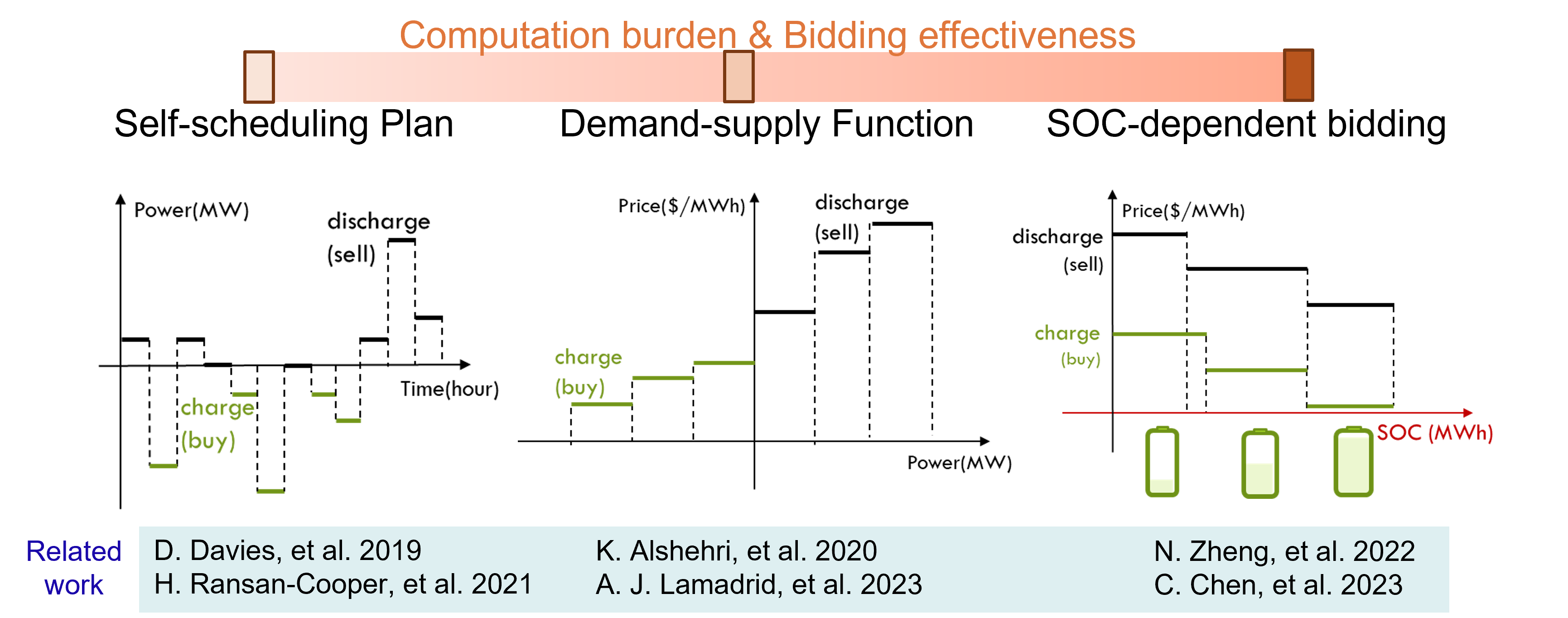}
	\caption{Three methods to encode batteries' opportunity costs. (From left to right we list diagrams and related works for the self-scheduling plan, the demand-supply function, and the SOC-dependent bidding, whose computation burden and bidding effectiveness increase sequentially.) }
	\label{fig:bidcompare}
\end{figure}

Many studies try to capture such opportunity costs of batteries in the electricity market, which in general fall into three categories (Fig.\ref{fig:bidcompare}). The first method is the self-scheduling plan, where batteries submit the optimal self-scheduling to the electricity market as non-dispatchable resources. Batteries typically optimize arbitrage opportunity profits with price forecasts to produce self-scheduling plans, which have less computation burdens but limited abilities in revealing the charging/discharging willingness of batteries under different prices \cite{davies2019combined, ransan2021applying, beuse2021comparing, qin2023role}. The second is the demand-supply curve (or demand-supply function),\footnote{Demand supply function is the inverse function of the demand-supply curve.} using price-quantity pairs to reveal batteries' willingness to charge and discharge \cite{JiangPowell15Storage, Lamadrid24JRE, Alshehri&etal:20TPS, attarha23TEMPR}. Both the self-scheduling and the demand-supply function are adopted by many electricity markets \cite{CAISO_StorageBid:21, ERCOT_storage:24}. The last is the SOC-dependent bidding from the recent proposal in California \cite{CAISO_SOCdependent:22}, allowing varying bid-in prices at different SOC levels. Although more effective in revealing the battery's power supply willingness, the SOC-dependent bidding hasn't been implemented because of the computation burden and the under-explored strategic behavior \cite{CAISO_DEB:20, ZhengXu22socAribitrage, ChenTong22arXivSoC}. Overall the demand-supply curve strikes a balance in the spectrum for computation complexity and effectiveness in revealing batteries' willingness. Therefore, we focus on the demand-supply curve in this paper and derive a simpler staircase compared to existing research relying on dynamic program \cite{JiangPowell15Storage, Lamadrid24JRE} and parametric program \cite{Alshehri&etal:20TPS, attarha23TEMPR}.

%% file: method_v1.tex
\section{Methods}

In this method section, we first provide the base case (ideal battery) and the proof of its demand-supply function (for simplicity). Then, we provide generalizations considering battery efficiencies, energy dissipation, and the ending SOC limit. Detailed proofs of the generalized demand-supply functions are provided in the supplemental material.

\subsection{Ideal battery's operation model used to generate the demand-supply function}
To generate the demand-supply function of a battery, the following multi-period look-ahead optimization model is adopted as our basic model:
\begin{equation}\label{eq:M1_1}
	M_1: \min_{\{P,E\}} \,\, J =\sum_{t=1}^{T}{-c_t P_t \Delta t}
\end{equation}	
subject to the following constraints for all $t \in [1,T]$
\begin{equation}\label{eq:M1_2}
	(\underline{\alpha}_t, \ \overline{\alpha}_t):\ -\overline{P} \leq P_t \leq \overline{P},
\end{equation}			
\begin{equation}\label{eq:M1_3}
	(\sigma_t):E_{t+1} = E_1 - \sum_{\tau=1}^{t} P_\tau  \Delta t,
\end{equation}	
\begin{equation}\label{eq:M1_4}
	(\underline{\beta}_{t+1}, \ \overline{\beta}_{t+1}):\ \underline{E} \leq E_{t+1} \leq \overline{E},
\end{equation}			
where $c_t$ is the electricity price at time $t$; $P_t$ is the net power output of the battery, for which positive means discharging and negative means charging. $E_t$ is the stored energy at time $t$ before the discharging and charging process at time $t$; $T$ is the number of time intervals in one day and $\Delta t$ is the time length of one single interval, i.e. $\Delta t = (24~\text{hour})/T$; $\overline{P}$ is the maximum discharging and charging power; $\overline{E}$ and $\underline{E}$ are the upper and lower limits of the stored energy; $E_1$ is the initial stored energy; $\underline{\alpha}_t$  and  $\overline{\alpha}_t$ are the multipliers of constraint \eqref{eq:M1_2}; $\sigma_t$ is the multiplier of constraint \eqref{eq:M1_3}; $\underline{\beta}_t$  and  $\overline{\beta}_t$ are the multipliers of constraint \eqref{eq:M1_4}. 

\par The objective in (\ref{eq:M1_1}) is the total cost of $T$ time periods. Constraints (\ref{eq:M1_2})  limit the upper and lower bounds of power;  constraint (\ref{eq:M1_3}) demonstrates the relationship between the stored energy at time $t+1$ considering the charging and discharging process before time $t+1$; constraint (\ref{eq:M1_4}) limits the upper and lower bounds of the stored energy.

\subsection{Results for Ideal Batteries}
The model in \eqref{eq:M1_1} is standard and widely adopted in the literature \cite{davies2019combined, ransan2021applying, beuse2021comparing, qin2023role, JiangPowell15Storage, Lamadrid24JRE, Alshehri&etal:20TPS, attarha23TEMPR}. Given a set of battery parameters, future price forecast $c_2, ...c_T$, and an initial SOC $E_1$, we solve the optimal charging and discharging power with different prices for the current interval $c_1$ and formulate $P_1$ as a function of $c_1$. Such a function $P_1=f(c_1)$ is exactly the demand-supply function studied in this paper.

We first establish the shape of the demand-supply function $P_1=f(c_1)$:
\newtheorem{lemma}{Lemma}
\begin{lemma}\label{BaseCaselemma3}
	\textit{Consider model $M_1$ and assume $\bar{P}>0$, $\bar{E}>\underline{E}$, then the demand-supply function of an ideal battery $P_1=f(c_1)$ solved from $M_1$ is staircase and monotonic non-decreasing.} 
\end{lemma}
\begin{proof}[Proof:]
    \par The dual of $M_1$ is also a linear programming, in the form of :
    \begin{equation}
        M_{1,\text{(dual)}}:  \min_{u} \, W = b'u 
    \end{equation}   
    subject to
    \begin{equation}
        (P): \, G'u \leq c 
    \end{equation}
    According to properties of the parametric linear programming problem (PLP) (Theorem 6.5 in \cite{borrelli2017predictive}), the optimal value function $W^*(c_1)$ is a piecewise affine and convex (with the price $c_1$ as a parameter). Thus $P_1=f(c_1)=\frac{\partial W^*}{\partial c_1}$ is a monotonic non-decreasing staircase function under the assumption of strong duality.
\end{proof}

We next investigate the physical background of these stairs. If the energy limit \eqref{eq:M1_4} never becomes binding, the optimal $P_1^*=\bar{P}$ (fully discharge) when $c_1\geq 0$ and $P_1^*=-\bar{P}$ (fully charge) when $c_1< 0$. These two stairs are more intuitive. To better understand stairs between them, we need to consider cases with binding energy limits for some periods, assuming the battery's capacity reaches its limit for the first time after changing or discharging in period $T_B$.

\begin{lemma}\label{BaseCaseN1less}	
 \textit{Consider model $M_1$ and assume there exists $T_B\leq T$ such that i) $E_{T_B+1}=\bar{E}$ or $\underline{E}$, and (ii) $\forall t\in[1, T_B], \underline{E}<E_t<\bar{E}$, further assume that $\forall t, t' \in [2, T_B], c_t\neq c_{t'} (t\neq t')$. Then the optimal solution to $M_1$ satisfy $T_B-n_c-n_d\leq1$, where $n_c$ and $n_d$ are numbers of fully charging and fully discharging periods till $T_B$, respectively.} 
\end{lemma}

\begin{proof}[Proof of Lemma~{\upshape\ref{BaseCaseN1less}}]

From the KKT condition of $M_1$, we have:
\begin{align}
    &\frac{\partial L}{\partial P_t} = -c_t \Delta t - \underline{\alpha_t} + \overline{\alpha_t} + \sum_{\tau\geq t}{\sigma_{\tau} \Delta t} = 0, \ \forall t \in [1,T]. \label{eq:omegaKKT_P} \\
    &\frac{\partial L}{\partial E_t} = \sigma_{t-1} -\underline{\beta}_t + \overline{\beta}_t= 0, \ \forall t \in [2,T+1]. \label{eq:omegaKKT_E}
\end{align}
For all $t\in[2, T_B]$, capacity limits \eqref{eq:M1_4} are inactive, thus $\sigma_{t-1}=\underline{\beta}_t =\overline{\beta}_t=0$ and $\sigma_{T_B}$ may become non-zero for the first time. Accordingly, $\sum_{\tau\geq t}{\sigma_{\tau}}$ share the same value for all $t\in[1, T_B]$, denoted by $\tilde{c} $. Substituting it to \eqref{eq:omegaKKT_P} we have

\begin{align}
    c_t \Delta t + \underline{\alpha_t} - \overline{\alpha_t} =  \tilde{c} \Delta t, \ \forall t \in [1,T_B], \label{eq:KKT1}
\end{align}
Equation \eqref{eq:KKT1} reveals the rationale of the battery's actions from period 1 to $T_B$ by comparing $c_t, t\in[1,T_B]$ to the threshold $\tilde{c}$: (i) If $c_t>\tilde{c}$, then $\overline{\alpha_t}>0$ the optimal action is to discharge fully; (ii) If $c_t<\tilde{c}$, then $\underline{\alpha_t}>0$ the optimal action is to charge fully; (iii) If the optimal action is neither fully charge nor fully discharge, there must be $c_t=\tilde{c}$. Periods fall into the last category are referred to as ``marginal periods''. Assuming all periods prior to $T_B$ have different prices, there is at most one marginal period and the conclusion of this lemma is true.

\end{proof}

Examples of the optimal serial of actions is illustrated in Fig.\ref{fig:figz}, which is consistent with our discussions below equation \eqref{eq:KKT1}. We can observe that similar phenomena appear between subsequent energy-limit binding periods. However, its proof is omitted in this article since it is less relevant. 

In some practical scenarios, the assumption of unequal prices over periods $[1, T]$ may not hold. However, we can always add $\epsilon$ to create distinct prices. Although the optimal solution to M1 may jump with minor price changes, its optimal value will not, nor will the demand-supply function. Usually, edges between stairs will shift $\epsilon$ to the left or right. When $\epsilon\rightarrow 0$, its impact on the curve also becomes negligible.

\begin{lemma}\label{BaseCaselemma2}
\textit{Under the unequal assumption as Lemma \ref{BaseCaseN1less}	and consider the price of the first period $c_1$ increases to $c_1' > c_1$ with all other parameters unchanged. Assume in both settings, the optimal actions in the first period are unique with $-\bar{P}<P_1^*, P_1'^*<\bar{P}$. Then $\exists T_B'\leq T$ being the first period where the capacity limit is binding with $c_1'$, and $E_{k+1}'\leq E_{k+1}, \forall k\leq \min\{T_B, T_B'\}$.}

\end{lemma}
\begin{proof}[Proof of Lemma~{\upshape\ref{BaseCaselemma2}}]

The existence of $T_B'$ is similar to that of $T_B$: Assuming $-\bar{P}<P_1'^*<\bar{P}$ and had energy limits never become active, changing discharge power at period 1 by $\epsilon$ (positive when $c_1 \geq 0$ and negative otherwise) is always better off. 

To prove the second item, given $k\leq\min\{T_B,T_B'\}$, we split the model $M_1$ into two parts as

\begin{equation}\label{eq:M2_1}
	M_2: \min_{E_{k+1}} \,\, J =J_{[1,k]} (E_{k+1}) + J_{[k+1,T]}(E_{k+1}),
\end{equation}	
where both terms on the right-hand side are functions of $E_{k+1}$ only. They represent optimal costs solved from $M_1$ for periods $[1,k]$ and $[k+1,T]$, respectively. For the former, an additional constraint specifying the energy after period $k$ is needed:

\begin{equation}\label{eq:M3_2}
	 (\eta) E_1-\sum_{t=1}^{k}{P_t \Delta t} = E_{k+1}.
\end{equation}	

For any optimal solution to $M_1$, its $E_{k+1}$ is also optimal for $M_2$, and vice versa.

We first look at the problem $[1,k]$. After adding the new constraint \eqref{eq:M3_2}, the original equation \eqref{eq:omegaKKT_P} for the first period becomes

\begin{equation}\label{eq:newP}
	 \frac{\partial L}{\partial P_1} = -c_1 \Delta t - \underline{\alpha_1} + \overline{\alpha_1} + \sum_{\tau\geq 1}^{k}{\sigma_{\tau} \Delta t} -\eta \Delta t= 0.
\end{equation}	

We have assumed that energy limits are never active $\forall t\in[1,k]$. From \eqref{eq:omegaKKT_E}, there is $\sigma_t, \underline{\beta}_{t+1}, \bar{\beta}_{t+1}=0, \forall t\in [1,k]$. Under the assumption of the first period being marginal, both $\underline{\alpha_1}$ and $\overline{\alpha_1}$ are equal to zero. According to the envelop theorem we then have

\begin{equation}\label{eq:M3_3}
	\nabla_{E_{k+1}} J_{[1,k]} (E_{k+1})=-\eta=c_1.
\end{equation}

For the unconstrained optimization $M_2$, its optimal solution yields 
\begin{equation}\label{eq:M3_4}
	\nabla_{E_{k+1}} J = \nabla_{E_{k+1}} J_{[1,k]} (E_{k+1}) + \nabla_{E_{k+1}}J_{[k+1,T]}(E_{k+1})=0.
\end{equation}
\begin{equation}\label{eq:M3_5}
	\nabla_{E_{k+1}}J_{[k+1,T]}(E_{k+1})= -\nabla_{E_{k+1}} J_{[1,k]} (E_{k+1}) =-c_1.
\end{equation}

Since $c_1$ is not a part of the problem $[k+1,T]$, it has no effect on the function $J_{[k+1,T]}(E_{k+1})$, which is characterized by a standard LP. According to the parametric linear programming theory, the function $J_{[k+1,T]}(E_{k+1})$ is piecewise affine and convex. When $c_1$ increases, indicating its derivative decreases, the corresponding $E_{k+1}$ must not increase due to the convexity of $J_{[k+1,T]}(E_{k+1})$. Thus $E_{k+1}'\leq E_{k+1}$.

\end{proof}

With all the lemmas above, we are now ready to establish the theorem for an ideal battery: 

\begin{theorem}\label{BaseCaseTHM_method}
    Under assumptions of lemma 1 and 2, the demand-supply function of an ideal battery is a staircase function with no more than five segments. With the increase of $c_1$, they appear in the following order: (1) Fully charge, (2) charge-for-charge (charge in the first period and reaching the energy upper bound in period $T_B$), (3) charge-for-discharge (reaching the energy lower bound in period $T_B$) or discharge-for-charge, (4) discharge-for-discharge, (5) fully discharge.
\end{theorem}

\begin{proof}[Proof of Theorem~{\upshape\ref{BaseCaseTHM_method}}]

We first discuss the existence of possible stairs and their physical backgrounds. Consider equation \eqref{eq:omegaKKT_P} for the first period, as long as $E_2$ does not reach its limits, there is $\sigma_1 = 0$ and  $\sum_{\tau\geq t}{\sigma_{\tau} \Delta t}$ is independent to the value of $c_1$. Thus if $c_1>\sum_{\tau\geq t}{\sigma_{\tau}}$, $\bar{\alpha_1}>0$, indicating the battery should fully discharge, and it should fully charge when  $c_1<\sum_{\tau\geq t}{\sigma_{\tau}}$.

For cases where the battery is neither fully charging nor fully discharging, it is safe to assume the existence of $T_B$ as the period after which the energy reaches its upper or lower limit for the first time. According to lemma 2, all periods in $[2,T_B]$ should either fully charge or fully discharge. Therefore we have

\begin{equation}\label{eq:Theorem1_1}
         P_{1} = \frac{E_1 - E_{T_B+1}}{\Delta t} - \Delta n\overline{P}.
\end{equation}
where $\Delta n$ is the difference between numbers of fully-discharge and fully-charge periods in $[2,T_B]$. Note that $E_{T_B+1}$ equals to either $\bar{E}$ or $\underline{E}$. Given the value of $E_{T_B+1}$ and the fact that $-\bar{P}<P_1<\bar{P}$, the integer $\Delta n$ can only pick from two consecutive values, denoted by $\nu$ and $\nu+1$. Their combinations define the four candidate segments in the middle of the demand-supply function:

\begin{itemize}
    \item \textbf{Charge-for-charge}: With $E_{T_B+1}=\bar{E}$ and $\Delta n=\nu+1$. Because $\nu\bar{P}\leq(E_1 - E_{T_B+1})/\Delta t\leq(\nu+1)\bar{P}$, there is $P_1\leq0$. Charge the remainder energy in the first period so that the battery reaches its upper energy bound after period $T_B$.
    \item \textbf{Charge-for-discharge}: With $E_{T_B+1}=\underline{E}$, $\Delta n=\nu+1$, and $P_1\leq0$. Charge the remainder energy in the first period so that the battery reaches its lower energy bound after period $T_B$.
    \item \textbf{Discharge-for-charge}: With $E_{T_B+1}=\bar{E}$, $\Delta n=\nu$, and $P_1\geq0$. 
    \item \textbf{Discharge-for-discharge}: With $E_{T_B+1}=\underline{E}$, $\Delta n=\nu$, and $P_1\geq0$. 
\end{itemize}
    
With the increase of $c_1$, lemma 1 proves that the discharge power $P_1$ never decreases, and lemma 3 proves that energy before reaching limits never increases. Therefore, if they exist, segments starting with ``charge'' must appear before those starting with ``discharge'', and segments ending with ``charge'' must appear before those ending with ``discharge''. Therefore, Charge-for-discharge and Discharge-for-charge can never appear together, and the order of all segments (if they exist) should follow the statement of this theorem.

\end{proof}

%% file: sn-article_v1.bbl
\begin{thebibliography}{10}
\expandafter\ifx\csname url\endcsname\relax
  \def\url#1{\burl{#1}}\fi
\expandafter\ifx\csname urlprefix\endcsname\relax\def\urlprefix{URL }\fi
\providecommand{\bibinfo}[2]{#2}
\providecommand{\eprint}[2][]{\url{#2}}
\providecommand{\doi}[1]{\url{https://doi.org/#1}}
\bibcommenthead

\bibitem{nationalacademies}
\bibinfo{title}{Accelerating decarbonization in the united states: Technology, policy, and societal dimensions}.
\newblock \bibinfo{howpublished}{[ONLINE], (2024/8/22) available at \url{https://www.nationalacademies.org/our-work/accelerating-decarbonization-in-the-united-states-technology-policy-and-societal-dimensions}} (\bibinfo{year}{2023}).

\bibitem{mckinsey}
\bibinfo{title}{The net-zero transition}.
\newblock \bibinfo{howpublished}{[ONLINE], (2024/8/22) available at \url{https://www.mckinsey.com/capabilities/sustainability/our-insights/the-net-zero-transition-what-it-would-cost-what-it-could-bring}} (\bibinfo{year}{2022}).

\bibitem{Chinaecologyandenvironment}
\bibinfo{title}{Progress report of china’s national carbon market}.
\newblock \bibinfo{howpublished}{[ONLINE], (2024/8/22) available at \url{https://www.mee.gov.cn/ywdt/xwfb/202407/W020240722528850763859.pdf}} (\bibinfo{year}{2024}).

\bibitem{iearenewables}
\bibinfo{title}{Renewables 2023}.
\newblock \bibinfo{howpublished}{[ONLINE], (2024/8/22) available at \url{https://www.iea.org/reports/renewables-2023}} (\bibinfo{year}{2024}).

\bibitem{verpoort2024impact}
\bibinfo{author}{Verpoort, P.~C.}, \bibinfo{author}{Gast, L.}, \bibinfo{author}{Hofmann, A.} \& \bibinfo{author}{Ueckerdt, F.}
\newblock \bibinfo{title}{Impact of global heterogeneity of renewable energy supply on heavy industrial production and green value chains}.
\newblock \emph{\bibinfo{journal}{Nature Energy}} \bibinfo{pages}{1--13} (\bibinfo{year}{2024}).

\bibitem{rode2021estimating}
\bibinfo{author}{Rode, A.} \emph{et~al.}
\newblock \bibinfo{title}{Estimating a social cost of carbon for global energy consumption}.
\newblock \emph{\bibinfo{journal}{Nature}} \textbf{\bibinfo{volume}{598}}, \bibinfo{pages}{308--314} (\bibinfo{year}{2021}).

\bibitem{millstein2021solar}
\bibinfo{author}{Millstein, D.} \emph{et~al.}
\newblock \bibinfo{title}{Solar and wind grid system value in the united states: The effect of transmission congestion, generation profiles, and curtailment}.
\newblock \emph{\bibinfo{journal}{Joule}} \textbf{\bibinfo{volume}{5}}, \bibinfo{pages}{1749--1775} (\bibinfo{year}{2021}).

\bibitem{USeia}
\bibinfo{title}{Solar and wind power curtailments are rising in california}.
\newblock \bibinfo{howpublished}{[ONLINE], (2024/8/22) available at \url{https://www.eia.gov/todayinenergy/detail.php?id=60822}} (\bibinfo{year}{2023}).

\bibitem{frew2021curtailment}
\bibinfo{author}{Frew, B.} \emph{et~al.}
\newblock \bibinfo{title}{The curtailment paradox in the transition to high solar power systems}.
\newblock \emph{\bibinfo{journal}{Joule}} \textbf{\bibinfo{volume}{5}}, \bibinfo{pages}{1143--1167} (\bibinfo{year}{2021}).

\bibitem{heptonstall2021systematic}
\bibinfo{author}{Heptonstall, P.~J.} \& \bibinfo{author}{Gross, R.~J.}
\newblock \bibinfo{title}{A systematic review of the costs and impacts of integrating variable renewables into power grids}.
\newblock \emph{\bibinfo{journal}{nature energy}} \textbf{\bibinfo{volume}{6}}, \bibinfo{pages}{72--83} (\bibinfo{year}{2021}).

\bibitem{degen2023energy}
\bibinfo{author}{Degen, F.}, \bibinfo{author}{Winter, M.}, \bibinfo{author}{Bendig, D.} \& \bibinfo{author}{T{\"u}bke, J.}
\newblock \bibinfo{title}{Energy consumption of current and future production of lithium-ion and post lithium-ion battery cells}.
\newblock \emph{\bibinfo{journal}{Nature energy}} \textbf{\bibinfo{volume}{8}}, \bibinfo{pages}{1284--1295} (\bibinfo{year}{2023}).

\bibitem{hunter2021techno}
\bibinfo{author}{Hunter, C.~A.} \emph{et~al.}
\newblock \bibinfo{title}{Techno-economic analysis of long-duration energy storage and flexible power generation technologies to support high-variable renewable energy grids}.
\newblock \emph{\bibinfo{journal}{Joule}} \textbf{\bibinfo{volume}{5}}, \bibinfo{pages}{2077--2101} (\bibinfo{year}{2021}).

\bibitem{sepulveda2021design}
\bibinfo{author}{Sepulveda, N.~A.}, \bibinfo{author}{Jenkins, J.~D.}, \bibinfo{author}{Edington, A.}, \bibinfo{author}{Mallapragada, D.~S.} \& \bibinfo{author}{Lester, R.~K.}
\newblock \bibinfo{title}{The design space for long-duration energy storage in decarbonized power systems}.
\newblock \emph{\bibinfo{journal}{Nature Energy}} \textbf{\bibinfo{volume}{6}}, \bibinfo{pages}{506--516} (\bibinfo{year}{2021}).

\bibitem{dowling2020role}
\bibinfo{author}{Dowling, J.~A.} \emph{et~al.}
\newblock \bibinfo{title}{Role of long-duration energy storage in variable renewable electricity systems}.
\newblock \emph{\bibinfo{journal}{Joule}} \textbf{\bibinfo{volume}{4}}, \bibinfo{pages}{1907--1928} (\bibinfo{year}{2020}).

\bibitem{wang2023accelerating}
\bibinfo{author}{Wang, Y.} \emph{et~al.}
\newblock \bibinfo{title}{Accelerating the energy transition towards photovoltaic and wind in china}.
\newblock \emph{\bibinfo{journal}{Nature}} \textbf{\bibinfo{volume}{619}}, \bibinfo{pages}{761--767} (\bibinfo{year}{2023}).

\bibitem{ferc}
\bibinfo{title}{Electric storage participation in markets operated by regional transmission organizations and independent system operators}.
\newblock \bibinfo{howpublished}{[ONLINE], (2024/8/22) available at \url{https://www.ferc.gov/media/order-no-841}} (\bibinfo{year}{2018}).

\bibitem{peng2023heterogeneous}
\bibinfo{author}{Peng, L.}, \bibinfo{author}{Mauzerall, D.~L.}, \bibinfo{author}{Zhong, Y.~D.} \& \bibinfo{author}{He, G.}
\newblock \bibinfo{title}{Heterogeneous effects of battery storage deployment strategies on decarbonization of provincial power systems in china}.
\newblock \emph{\bibinfo{journal}{Nature communications}} \textbf{\bibinfo{volume}{14}}, \bibinfo{pages}{4858} (\bibinfo{year}{2023}).

\bibitem{comello2019emergence}
\bibinfo{author}{Comello, S.} \& \bibinfo{author}{Reichelstein, S.}
\newblock \bibinfo{title}{The emergence of cost effective battery storage}.
\newblock \emph{\bibinfo{journal}{Nature communications}} \textbf{\bibinfo{volume}{10}}, \bibinfo{pages}{2038} (\bibinfo{year}{2019}).

\bibitem{he2018intertemporal}
\bibinfo{author}{He, G.}, \bibinfo{author}{Chen, Q.}, \bibinfo{author}{Moutis, P.}, \bibinfo{author}{Kar, S.} \& \bibinfo{author}{Whitacre, J.~F.}
\newblock \bibinfo{title}{An intertemporal decision framework for electrochemical energy storage management}.
\newblock \emph{\bibinfo{journal}{Nature Energy}} \textbf{\bibinfo{volume}{3}}, \bibinfo{pages}{404--412} (\bibinfo{year}{2018}).

\bibitem{levin2023energy}
\bibinfo{author}{Levin, T.} \emph{et~al.}
\newblock \bibinfo{title}{Energy storage solutions to decarbonize electricity through enhanced capacity expansion modelling}.
\newblock \emph{\bibinfo{journal}{Nature Energy}} \textbf{\bibinfo{volume}{8}}, \bibinfo{pages}{1199--1208} (\bibinfo{year}{2023}).

\bibitem{davies2019combined}
\bibinfo{author}{Davies, D.} \emph{et~al.}
\newblock \bibinfo{title}{Combined economic and technological evaluation of battery energy storage for grid applications}.
\newblock \emph{\bibinfo{journal}{Nature Energy}} \textbf{\bibinfo{volume}{4}}, \bibinfo{pages}{42--50} (\bibinfo{year}{2019}).

\bibitem{ransan2021applying}
\bibinfo{author}{Ransan-Cooper, H.}, \bibinfo{author}{Sturmberg, B.~C.}, \bibinfo{author}{Shaw, M.~E.} \& \bibinfo{author}{Blackhall, L.}
\newblock \bibinfo{title}{Applying responsible algorithm design to neighbourhood-scale batteries in australia}.
\newblock \emph{\bibinfo{journal}{Nature Energy}} \textbf{\bibinfo{volume}{6}}, \bibinfo{pages}{815--823} (\bibinfo{year}{2021}).

\bibitem{beuse2021comparing}
\bibinfo{author}{Beuse, M.}, \bibinfo{author}{Steffen, B.}, \bibinfo{author}{Dirksmeier, M.} \& \bibinfo{author}{Schmidt, T.~S.}
\newblock \bibinfo{title}{Comparing co2 emissions impacts of electricity storage across applications and energy systems}.
\newblock \emph{\bibinfo{journal}{Joule}} \textbf{\bibinfo{volume}{5}}, \bibinfo{pages}{1501--1520} (\bibinfo{year}{2021}).

\bibitem{qin2023role}
\bibinfo{author}{Qin, X.}, \bibinfo{author}{Xu, B.}, \bibinfo{author}{Lestas, I.}, \bibinfo{author}{Guo, Y.} \& \bibinfo{author}{Sun, H.}
\newblock \bibinfo{title}{The role of electricity market design for energy storage in cost-efficient decarbonization}.
\newblock \emph{\bibinfo{journal}{Joule}} \textbf{\bibinfo{volume}{7}}, \bibinfo{pages}{1227--1240} (\bibinfo{year}{2023}).

\bibitem{JiangPowell15Storage}
\bibinfo{author}{Jiang, D.~R.} \& \bibinfo{author}{Powell, W.~B.}
\newblock \bibinfo{title}{Optimal hour-ahead bidding in the real-time electricity market with battery storage using approximate dynamic programming}.
\newblock \emph{\bibinfo{journal}{INFORMS Journal on Computing}} \textbf{\bibinfo{volume}{27}}, \bibinfo{pages}{525--543} (\bibinfo{year}{2015}).
\newblock \urlprefix\url{https://doi.org/10.1287/ijoc.2015.0640}.

\bibitem{Lamadrid24JRE}
\bibinfo{author}{Lamadrid, A.~J.}, \bibinfo{author}{Lu, H.} \& \bibinfo{author}{Mount, T.~D.}
\newblock \bibinfo{title}{A simple way to integrate distributed storage into a wholesale electricity market}.
\newblock \emph{\bibinfo{journal}{Journal of Regulatory Economics}} \textbf{\bibinfo{volume}{65}}, \bibinfo{pages}{27--63} (\bibinfo{year}{2024}).

\bibitem{Alshehri&etal:20TPS}
\bibinfo{author}{Alshehri, K.}, \bibinfo{author}{Ndrio, M.}, \bibinfo{author}{Bose, S.} \& \bibinfo{author}{Ba{\c{s}}ar, T.}
\newblock \bibinfo{title}{Quantifying market efficiency impacts of aggregated distributed energy resources}.
\newblock \emph{\bibinfo{journal}{IEEE Transactions on Power Systems}} \textbf{\bibinfo{volume}{35}}, \bibinfo{pages}{4067--4077} (\bibinfo{year}{2020}).

\bibitem{attarha23TEMPR}
\bibinfo{author}{Attarha, A.} \emph{et~al.}
\newblock \bibinfo{title}{Adjustable price-sensitive {DER} bidding within network envelopes}.
\newblock \emph{\bibinfo{journal}{IEEE Transactions on Energy Markets, Policy and Regulation}} \textbf{\bibinfo{volume}{1}}, \bibinfo{pages}{248--258} (\bibinfo{year}{2023}).

\bibitem{CAISO_StorageBid:21}
\bibinfo{title}{Energy storage enhancements straw proposal}.
\newblock \bibinfo{howpublished}{[ONLINE], available (2024/8/11) at \url{https://stakeholdercenter.caiso.com/InitiativeDocuments/Presentation-EnergyStorageEnhancements-Dec14-2021.pdf}} (\bibinfo{year}{2021}).

\bibitem{ERCOT_storage:24}
\bibinfo{title}{{ERCOT:} battery energy storage}.
\newblock \bibinfo{howpublished}{[ONLINE], available (2024/8/11) at \url{https://www.ercot.com/mktrules/keypriorities/bes}} (\bibinfo{year}{2019}).

\bibitem{CAISO_SOCdependent:22}
\bibinfo{title}{Energy storage enhancements revised straw proposal}.
\newblock \bibinfo{howpublished}{[ONLINE], available (2024/8/22) at \url{http://www.caiso.com/InitiativeDocuments/RevisedStrawProposal-EnergyStorageEnhancements.pdf}} (\bibinfo{year}{2022}).

\bibitem{CAISO_DEB:20}
\bibinfo{title}{Energy storage and distributed energy resources – storage default energy bid}.
\newblock \bibinfo{howpublished}{[ONLINE], available (2024/8/22) at \url{http://www.caiso.com/InitiativeDocuments/FinalProposal-EnergyStorage-DistributedEnergyResourcesPhase4-DefaultEnergyBid.pdf}} (\bibinfo{year}{2020}).

\bibitem{ZhengXu22socAribitrage}
\bibinfo{author}{Zheng, N.}, \bibinfo{author}{Jaworski, J.~J.} \& \bibinfo{author}{Xu, B.}
\newblock \bibinfo{title}{Arbitraging variable efficiency energy storage using analytical stochastic dynamic programming}.
\newblock \emph{\bibinfo{journal}{IEEE Transactions on Power Systems}} \bibinfo{pages}{1--1} (\bibinfo{year}{2022}).

\bibitem{ChenTong22arXivSoC}
\bibinfo{author}{Chen, C.} \& \bibinfo{author}{Tong, L.}
\newblock \bibinfo{title}{Convexifying market clearing of {SoC-Dependent} bids from merchant storage participants}.
\newblock \emph{\bibinfo{journal}{IEEE Transactions on Power Systems}} \textbf{\bibinfo{volume}{38}}, \bibinfo{pages}{2955--2957} (\bibinfo{year}{2023}).

\bibitem{PJMLMPs}
\bibinfo{title}{Pjm lmps}.
\newblock \bibinfo{howpublished}{[ONLINE], available (2024/8/22) at \url{https://www.pjm.com/markets-and-operations/data-dictionary.aspx}} (\bibinfo{year}{2022}).

\bibitem{SZpricedata}
\bibinfo{title}{Shenzhen electricity price data}.
\newblock \bibinfo{howpublished}{[ONLINE], available (2024/8/22) at \url{https://fgw.sz.gov.cn/gkmlpt/content/9/9493/post_9493596.html\#2659}} (\bibinfo{year}{2021}).

\bibitem{borrelli2017predictive}
\bibinfo{author}{Borrelli, F.}, \bibinfo{author}{Bemporad, A.} \& \bibinfo{author}{Morari, M.}
\newblock \emph{\bibinfo{title}{Predictive control for linear and hybrid systems}}  (\bibinfo{publisher}{Cambridge University Press}, \bibinfo{year}{2017}).

\end{thebibliography}
